\pdfoutput=1
\documentclass[aps,prl,twocolumn,superscriptaddress,showpacs,preprintnumbers]{revtex4-1}
\usepackage[colorlinks=true, pdfstartview=FitV, linkcolor=red, citecolor=blue, urlcolor=black, pdftitle={Inhomogeneous Polyakov loop induced by inhomogeneous chiral condensates},pdfauthor={Tomoya Hayata, Arata Yamamoto},pdfsubject={}, pdfkeywords={}]{hyperref}

\usepackage[dvipdfmx]{graphicx}
\usepackage{times}
\usepackage{amssymb,bm}
\usepackage{amsthm,amsmath}
\usepackage{color}
\newcommand{\cL}{{\cal L}}
\newcommand{\Nf}{N_f}
\newcommand{\Nc}{N_c}
\newcommand{\sn}{{\rm sn}}
\newcommand{\cn}{{\rm cn}}
\newcommand{\dn}{{\rm dn}}
\newcommand{\tr}{{\rm tr}}

\begin{document}
\preprint{RIKEN-QHP-160}
\title{Inhomogeneous Polyakov loop induced by inhomogeneous chiral condensates}
\author{Tomoya Hayata}
\affiliation{Department of Physics, The University of Tokyo, Tokyo 113-0031, Japan}
\affiliation{Theoretical Research Division, Nishina Center, RIKEN, Wako 351-0198, Japan}
\author{Arata Yamamoto}
\affiliation{Department of Physics, The University of Tokyo, Tokyo 113-0031, Japan}
\affiliation{Theoretical Research Division, Nishina Center, RIKEN, Wako 351-0198, Japan}

\date{\today}

\begin{abstract}
We study the spatial inhomogeneity of the Polyakov loop induced by inhomogeneous chiral condensates.
We formulate an effective model of gluons on the background fields of chiral condensates, and perform its lattice simulation.
On the background of inhomogeneous chiral condensates, the Polyakov loop exhibits an in-phase spatial oscillation with the chiral condensates.
We also analyze the heavy quark potential and show that the inhomogeneous Polyakov loop indicates the inhomogeneous confinement of heavy quarks.
\end{abstract}
\maketitle

\paragraph{Introduction.}
Quantum chromodynamics (QCD), which describes dynamics of quarks and gluons, is the SU($\Nc$) gauge theory coupled with $N_f$-flavor fermions. 
The complexity of the interaction and the internal degrees of freedom enriches the phase structure of QCD.
There exist two essential phenomena characterizing the phase structure: spontaneous chiral symmetry breaking and color confinement.
The phase structure of these phenomena has been actively investigated as a function of various external parameters, such as temperature, quark number chemical potential (or other types of chemical potentials), magnetic field, and so on \cite{reviews}.

At finite density with the population imbalance between different quark species, there appears an inhomogeneous chiral phase~\cite{Deryagin:1992rw,Park:1999bz,Basar:2008im,Nickel:2009ke,Kojo:2009ha,Basar:2010zd,Buballa:2014tba}, which is an analogue to an inhomogeneous superconducting phase \cite{Fulde:1964zz}.
In the inhomogeneous chiral phase, translational invariance is spontaneously broken and the chiral condensate, which is an order parameter of spontaneous chiral symmetry breaking, spatially oscillates.
The existence of the inhomogeneous chiral phase is considered to affect the equation of state and transport properties of dense quark systems such as neutron stars.

The confinement-deconfinement phase transition is defined as spontaneous breaking of center symmetry $Z_{\Nc}$.
This symmetry is exact in the pure gauge sector.
The order parameter is the Polyakov loop, of which the expectation value is interpreted as the free energy of an infinitely heavy quark \cite{Polyakov:1978vu}.
Although $Z_{\Nc}$ symmetry is explicitly broken by quarks, the Polyakov loop plays a role of an approximate order parameter of the confinement even in full QCD.
 The susceptibility of the Polyakov loop no longer diverges but has a peak near the chiral phase transition point, and thus it suggests the sudden liberation of many degrees of freedom at this point~\cite{Cheng:2007jq}.

 While the inhomogeneity of chiral symmetry breaking has been investigated in details,
 the inhomogeneity of the confinement is far less known. 
 The inhomogeneous chiral condensate is induced by the mismatch between the Fermi surfaces of quarks.
 On the other hand, there is no counter part of the Fermi surface of gluons.
 However, since the deconfinement transition is entangled with chiral symmetry restoration~\cite{Hatta:2003ga,Fukushima:2003fw}, 
 it is natural that the Polyakov loop also becomes inhomogeneous through the coupling with the inhomogeneous chiral condensate.
 The study for the inhomogeneity of gauge fields is a highly nontrivial challenge, which gives us a new insight on the vacuum of gauge theories.

In this Letter, we discuss the spatial inhomogeneity of the Polyakov loop in inhomogeneous chiral phases.
Since it is terribly difficult to simulate inhomogeneous chiral condensates in full QCD due to the sign problem and the fine-tuning problem~\cite{Yamamoto:2014lia}, we study an effective model of gluons with the background chiral fields.
We first consider a chiral effective model which approximately reproduces the effect of chiral fields on the Polyakov loop, replace the chiral fields to the corresponding condensates, and then add them to the pure Yang-Mills action.
More specifically, we adopt the simplest forms of a glueball-meson interaction term and an explicit $Z_{\Nc}$ symmetry breaking term.
Using this effective model, we calculate the Polyakov loop and the heavy quark potential by lattice simulations.
We also discuss how different types of inhomogeneous chiral condensates do or do not affect the Polyakov loop.

\paragraph{Effective model.}
We consider a simple model to study the Polyakov loop on the background of chiral condensates.
The model Lagrangian is
\begin{equation}
\cL = \cL_{\rm YM} + \cL_{\sigma\pi} + \cL_{\rm P}.
\end{equation}
The first term is the SU($\Nc$) Yang-Mills Lagrangian:
\begin{equation}
\cL_{\rm YM} = \frac{1}{2g^2} \tr\; G_{\mu\nu}G^{\mu\nu},
\end{equation}
and the other terms are contributions of quarks.
The model is formulated so as to have the same symmetry as massless QCD at finite temperature.

The second term is the interaction between scalar and pseudoscalar mesons, and glueballs.
In this Letter, we consider the two flavor case, in which order parameters of chiral symmetry breaking are the chiral condensate $\langle\bar{q}{q}\rangle$ and the pion condensate $\langle\bar{q}i\gamma^5\tau^a{q}\rangle$.
We first consider a chiral effective theory of dynamical scalar and pseudoscalar mesons, $\sigma$ and $\pi^a$, which directly couple to gluons, and then replace the meson fields by classical background of the condensates $\sigma=\langle\bar{q}{q}\rangle$ and $\pi^a= \langle\bar{q}i\gamma^5\tau^a{q}\rangle$.
In the chiral limit, SU$_L(2)\times$SU$_R(2)$ chiral symmetry restricts the meson fields to the chiral symmetric combination $\Sigma^2 = \langle\bar{q}{q}\rangle^2 + \sum_a \langle\bar{q}i\gamma^5\tau^a{q}\rangle^2$.
The effective interaction is given as 
\begin{equation}
\cL_{\sigma\pi} = \left( c_0-c_2 \Sigma^2 \right) \tr\; G_{\mu\nu}G^{\mu\nu} ,
\label{eq:mixing}
\end{equation}
up to the quadratic order in the meson fields and the gluon field strength.
The zeroth order term means that quark loops reduce the gauge coupling constant.
The background chiral condensate induces an external potential through the interaction between glueballs and mesons, which leads to the interplay between chiral and deconfinement phase transitions~\cite{Hatta:2003ga}.
If the current quark mass is nonzero, it explicitly breaks chiral symmetry and thus induces the additional glueball-sigma mixing term $-c_1\sigma \tr G_{\mu\nu}G^{\mu\nu}$ in Eq.~\eqref{eq:mixing}. 

The third term is the Polyakov loop term.
Let us consider the effect of dynamical quarks in the fundamental representation of SU($\Nc$), which are not invariant under $Z_{\Nc}$ symmetry.
This effect can be taken into account as the background field of the Polyakov loop, which explicitly breaks $Z_{\Nc}$ symmetry.
The simplest form is
\begin{equation}
\cL_{\rm P} = -c_{\rm P} \left( e^{\mu/T} \tr L+e^{-\mu/T} \tr L^{\dagger} \right) .
\label{eq:z3breaking}
\end{equation}
where $L$ is the Polyakov loop.
The simplest way to estimate the coefficient $c_{\rm P}$ is to expand the quark thermodynamic potential: $\Omega/(VT) \sim\sum_{\epsilon} \{ \tr\ln[1+Ln]+\tr\ln[1+L^{\dagger}\bar{n}] \}$ with $n\; (\bar{n})=\exp(-(\epsilon\mp\mu)/T)$ being the thermal distribution of quarks (anti-quarks) \cite{Dumitru:2005ng}.
At the leading order,
\begin{equation}
c_{\rm P} = \frac{2 \Nf T}{(2\pi)^3} \int d^3 p \exp (-\sqrt{p^2+M^2}/T).
\label{eq:cp}
\end{equation}
The magnitude of $Z_{\Nc}$ symmetry breaking depends on the constituent quark mass $M$.
(The constituent quark mass is proportional to the chiral condensate, $M^2 \propto \Sigma^2$, in chiral effective models.)
It becomes larger as the constituent quark mass becomes smaller.
Therefore, when the chiral condensate varies with spatial position, the Polyakov loop can be inhomogeneous.

In the above formulation, we considered only the lowest order terms of operators which are necessary and sufficient for the following discussion.
We can also add higher-order terms, such as $\Sigma^4$ and $\tr L \tr L^\dagger$, or derivative terms, such as $ (\partial_\mu \langle\bar{q}{q}\rangle)^2 + \sum_a (\partial_\mu \langle\bar{q}i\gamma^5\tau^a{q}\rangle)^2$. 
In the following discussion, we drop these terms to keep our model as simple as possible.
Even if we add these terms, the qualitative conclusion does not change. 

\paragraph{Inhomogeneous chiral condensates.}

So far, several types of inhomogeneous chiral condensates have been proposed by using chiral effective models.
We restrict ourselves to one-dimensional modulation and take the $z$ axis as a direction of the modulation.

The first type of the modulation is a plane wave, which is an analogue to a spin density wave in condensed matter physics and called the chiral spiral or the dual chiral density wave~\cite{Park:1999bz,Nickel:2009ke}.
It is given explicitly as
\begin{align}
& \langle\bar{q}{q}\rangle(z)=\Sigma \cos(k z) , 
\notag \\
& \langle\bar{q}i\gamma^5\tau^3{q}\rangle(z)=\Sigma \sin(k z) .
\label{eq:spiral}
\end{align}
We take charged pion condensates zero because they are considered to be zero in any cases at zero isospin chemical potential.

The chiral spiral cannot induce any spatial inhomogeneity of gluons in the chiral limit.
Since the potential terms depend only on $\Sigma^2$ and $M^2$, the phase modulation cancels out and thus the potentials are homogeneous (except for the explicit chiral symmetry breaking term).
This is true even if we take into account higher-order or derivative terms.
This can be explained by the fact that the potentials can always be mapped to spatially uniform ones by a coordinate dependent transformation $\Phi \equiv \sigma + i \pi^3 \rightarrow \Phi^{\prime}=\exp(-ikz)\Phi={\rm const}.$
In other words, chiral symmetry is uniformly broken in the chiral spiral state in the sense that the amplitude of the condensates is constant, so that the deconfinement transition also uniformly occurs.

The second type is the real kink crystal condensate~\cite{Basar:2008im,Nickel:2009ke}, which is considered to be the most energetically favorable state in ($1+3$)-dimensional chiral effective models.
It is given as (in the chiral limit)
\begin{equation}
\langle\bar{q}{q}\rangle(z)=\nu\Delta\frac{\sn(\Delta z;\nu)\cn(\Delta z;\nu)}{\dn(\Delta z;\nu)} ,
\label{eq:kink}
\end{equation}
where $0\leq\nu\leq1$, and $\Delta$ is a real parameter. 
$\sn$, $\cn$ and $\dn$ are the Jacobi elliptic functions.
(For details, see Refs.~\cite{Basar:2008im,Nickel:2009ke}.)
In this state, only the scalar condensate is nonzero and the pseudoscalar condensate is zero.
There exists only the amplitude modulation  shown in Eq.~\eqref{eq:kink}, which gives a periodic array of kink and anti-kink modulations. 

The real kink crystal condensate can induce the amplitude modulation of the Polyakov loop.
Since $\Sigma^2$ and $M^2$ are inhomogeneous, chiral symmetry breaking is really inhomogeneous.
The effective interaction $\cL_{\sigma\pi}$ gives a periodic potential which modulates the local gauge coupling constant. 
Also, the Polyakov loop term $\cL_{\rm P}$ gives a periodic potential.
These modulation of the local gauge coupling constant and the magnitude of $Z_{\Nc}$ breaking do not vanish even in the chiral limit, so that the amplitude of the Polyakov loop may also modulate in the same period given by kink crystals.

Before conducting numerical simulations, we discuss the possibility that the phase modulation of the Polyakov loop is induced by inhomogeneous chiral condensates.
In our formulation, we need the interaction between glueballs and condensed mesons which makes the Polyakov loop favor the complex direction.
In fact, such an interaction can be induced via the axial anomaly. 
Due to the axial anomaly, U$_{A}(1)$ pseudoscalar meson condensate can induce a potential:
\begin{equation}
\cL_{\eta} =  c_{\eta} \langle\bar{q}i\gamma^5{q}\rangle \tr G_{\mu\nu}\widetilde{G}^{\mu\nu} ,
\end{equation}
with $\widetilde{G}^{\mu\nu}=i\epsilon^{\mu\nu\rho\sigma}G_{\rho\sigma}/2$.
The coefficient $c_{\eta}$ should be determined to reproduce the axial anomaly equation.
The U$_{A}(1)$ condensate plays the same role as the $\theta$ angle.
Since the $\theta$ angle may affect the Polyakov loop in a similar way of the imaginary chemical potential~\cite{Roberge:1986mm} as conjectured in Ref.~\cite{Bonati:2013tt}, the inhomogeneous U$_{A}(1)$ condensate may induce the phase modulation of the Polyakov loop near the deconfinement temperature.
In contrast, if the temperature is low enough, because of the Roberge-Weiss-like phase transition at finite $T$ and $\theta$~\cite{Bonati:2013tt}, the phase of Polyakov loop may jump as the U$_{A}(1)$ condensate crosses the critical value appearing in the period of $2\pi$.

Such a meson condensate appears in the quarkyonic chiral spiral state~\cite{Kojo:2009ha} and in the chiral magnetic spiral state~\cite{Basar:2010zd}, which are considered to be a possible ground state in dense quark system with large number of colors and in the presence of strong magnetic fields, respectively.
In the Monte Carlo simulation, however, this potential makes the effective gluonic action complex and causes the so-called sign problem even in our simple model.
We do not consider this potential in the following numerical simulations.

\paragraph{Numerical simulation.}

We study this model in lattice simulations.
Since this model is pure gauge theory with the Polyakov loop term, the numerical simulation can be done in the standard way of the SU(3) lattice gauge theory.
As a consequence of the interaction term, the lattice gauge coupling constant changes as
\begin{equation}
 \beta(\Sigma) = 4N_c\left( \frac{1}{2g^2} + c_0 - c_2 \Sigma^2 \right).
\end{equation}
When the chiral condensate is inhomogeneous, the lattice gauge coupling constant depends on spatial coordinate \cite{Huang:1990jf,Yamamoto:2013zwa}.
The field strength is evaluated as the symmetric average of plaquettes \cite{Yamamoto:2013zwa}.
The lattice volume is $N_xN_yN_z\times N_\tau = 12^3\times 6$ with periodic boundary conditions.
Since the chemical potential causes the sign problem in Eq.~\eqref{eq:z3breaking}, we set $\mu=0$, namely, $\langle\tr L\rangle=\langle\tr L^{\dagger}\rangle$.

In principle, the parameters, $c_0$, $c_2$, and $c_{\rm P}$, can be nonperturbatively determined by matching this model calculation with the full QCD simulation.
In this Letter, we naively evaluate these parameters as follows.
The parameter $c_0$ can be absorbed into the redefinition of the gauge coupling constant.
The parameter $c_2$ is estimated by matching the temperature in pure Yang-Mills theory with those in two-flavor QCD.
The phase transition temperature is $T_c \simeq 270$ MeV in pure Yang-Mills theory and $T_c \simeq 170$ MeV in two-flavor QCD.
To reproduce two-flavor QCD by this pure gauge model, the $\Sigma$-dependence of $\beta$ should change the phase transition temperature.
There should be two regions: $\langle\tr L\rangle=0$ and $\Sigma \ne 0$ in $T < T_c$, and $\langle\tr L\rangle \ne 0$ and $\Sigma = 0$ in $T > T_c$.
To connect these two regions at $T=T_c$, the necessary condition is $T(\Sigma = 0)/T(\Sigma \ne 0) > 270/170$ in the unit of pure Yang-Mills theory.
Based on the lattice unit in pure Yang-Mills simulations \cite{Takahashi:2002bw}, we set $\beta(\Sigma \ne 0) = 5.7$ and $\beta(\Sigma = 0) = 6.0$, which satisfy the above condition.
The parameter $c_{\rm P}$ is determined by Eq.~\eqref{eq:cp}.
The constituent quark mass is set at $M(\Sigma \ne 0) \simeq 0.3$ GeV and $M(\Sigma = 0) = 0$ GeV.

\begin{figure}[h]
\includegraphics[scale=1.3]{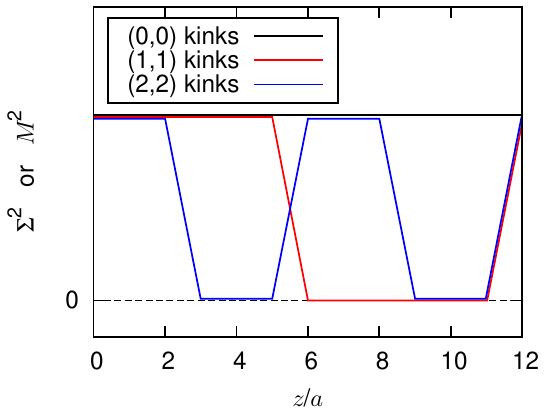}
\caption{\label{figL1}
Background chiral condensate and constituent quark mass.
The notation ``$(i,j)$ kinks'' denotes $i$ kinks and $j$ anti-kinks.
}
\end{figure}

\begin{figure}[h]
\includegraphics[scale=1.3]{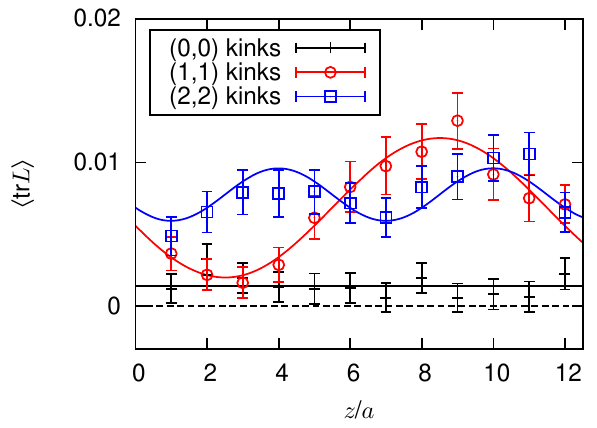}
\caption{\label{figL2}
The Polyakov loop on the backgrounds in Fig.~\ref{figL1}.
The solid curves are drawn for guide for eyes.
}
\end{figure}

We consider the backgrounds of chiral condensate and constituent quark mass shown in Fig.~\ref{figL1}, which is motivated by the real kink crystal condensate.
In Fig.~\ref{figL2}, we show the expectation value of the Polyakov loop on the backgrounds.
The Polyakov loop is nonzero even on the $(0,0)$-kink background because $Z_{\Nc}$ symmetry is explicitly broken.
The inhomogeneity of the chiral condensate affects the Polyakov loop.
The Polyakov loop decreases in the region where chiral symmetry is broken, and increases in the region where chiral symmetry is restored.
Thus, the modulation of the chiral condensate induces the amplitude modulation of the Polyakov loop, as clearly seen in Fig.~\ref{figL2}.

\begin{figure}[h]
\includegraphics[scale=1.3]{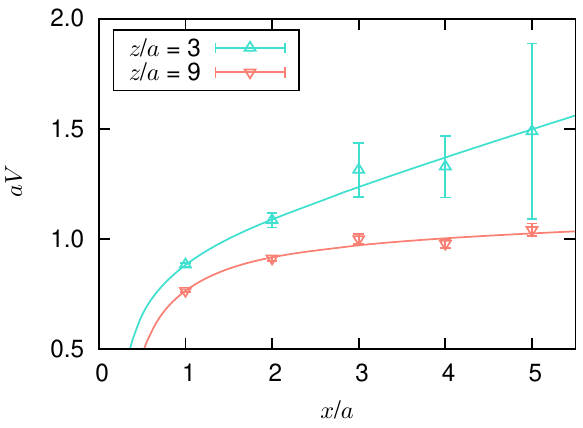}
\caption{\label{figL3}
Color averaged potential.
The data in the confinement domain ($z/a=3$) and in the deconfinement domain ($z/a=9$) of the $(2,2)$-kink background are shown.
}
\end{figure}

When $Z_{\Nc}$ symmetry is explicitly broken, the Polyakov loop is nothing more than an approximate order parameter.
To read more rigid information for the heavy quark confinement, we calculated the potential between a Polyakov loop and an anti-Polyakov loop:
\begin{equation}
V(x) = - T  \ln \langle \tr L(0) \tr L^\dagger (x)\rangle .
\end{equation}
This potential is called the color averaged potential because it is the average of the color singlet and octet potentials \cite{McLerran:1980pk}.
In Fig.~\ref{figL3}, we show the potential in the perpendicular direction to the modulation at fixed-$z$ planes.
We show two data of the $(2,2)$-kink background shown in Fig.~\ref{figL1}.
In the region where the Polyakov loop is suppressed ($z/a=3$), the potential includes the confining potential, and in the region where the Polyakov loop is enhanced ($z/a=9$), the confinement potential disappears.
Both of the confinement domain and the deconfinement domain exist locally.
Therefore the inhomogeneous Polyakov loop can be interpreted as inhomogeneous vacuum of the heavy quark confinement.
We note that the effect of string breaking is not considered in this pure gauge simulation.

 \paragraph{Concluding remarks.}
In this Letter, we have investigated the spatial inhomogeneity of the Polyakov loop on the background of inhomogeneous chiral condensates. 
We have numerically showed that the amplitude modulation of the Polyakov loop is induced by the real kink crystal condensate, which is the amplitude modulation of the chiral condensate.
The inhomogeneity of the Polyakov loop can be interpreted as the inhomogeneous confinement of heavy quarks, which means the generation of the local confinement and deconfinement domains.
Our model suggests that the phase modulation of the Polyakov loop can be induced by the axial anomaly.

We have constructed the effective model to study the effect of chiral condensates on dynamics of gluons.
The effect of quarks is reduced to external potentials, and the resultant model is described only by gluons.
In general, the lattice simulation with quarks is numerically more expensive than that of pure gauge theory.
Moreover, it sometimes suffers from the sign problem.
By introducing quarks as background fields, we can study dynamics of gluons effectively, though approximately.

 There are several future directions to investigate the inhomogeneous Polyakov loop.
 First, we need to improve our analysis quantitatively and discuss in which regions this state appears in the real QCD phase diagram.
 The analysis based on effective models of QCD, such as chiral effective models with the the Polyakov loop 
 or the gauge-gravity correspondence, is another interesting direction.
 It is also important to discuss phenomenological impacts of the inhomogeneous confinement, e.g., on heavy quarkonia in medium.

\begin{acknowledgements}
 We thank K.~Fukushima and Y.~Hidaka for stimulating discussions and useful comments.
 T.~H.~was supported by JSPS Research Fellowships for Young Scientists.
The numerical simulations were performed by using the RIKEN Integrated Cluster of Clusters (RICC) facility.
\end{acknowledgements}



\begin{thebibliography}{100}

  \bibitem{reviews} 
  For recent reviews,
  M.~G.~Alford, A.~Schmitt, K.~Rajagopal and T.~Sch\"afer,
  Rev.\ Mod.\ Phys.\  {\bf 80}, 1455 (2008)
  [arXiv:0709.4635 [hep-ph]];
  C.~DeTar and U.~M.~Heller,
  Eur.\ Phys.\ J.\ A {\bf 41}, 405 (2009)
  [arXiv:0905.2949 [hep-lat]];
  K.~Fukushima and T.~Hatsuda,
  Rept.\ Prog.\ Phys.\  {\bf 74}, 014001 (2011)
  [arXiv:1005.4814 [hep-ph]];
  D.~E.~Kharzeev, K.~Landsteiner, A.~Schmitt and H.~-U.~Yee,
  Lect.\ Notes Phys.\  {\bf 871}, 1 (2013)
  [arXiv:1211.6245 [hep-ph]].

\bibitem{Deryagin:1992rw} 
  D.~V.~Deryagin, D.~Y.~.Grigoriev and V.~A.~Rubakov,
  Int.\ J.\ Mod.\ Phys.\ A {\bf 7}, 659 (1992).
  
  \bibitem{Park:1999bz} 
  B.~-Y.~Park, M.~Rho, A.~Wirzba and I.~Zahed,
  Phys.\ Rev.\ D {\bf 62}, 034015 (2000)
  [hep-ph/9910347];
  V.~Schon and M.~Thies,
  Phys.\ Rev.\ D {\bf 62}, 096002 (2000)
  [hep-th/0003195];
  R.~Rapp, E.~V.~Shuryak and I.~Zahed,
  Phys.\ Rev.\ D {\bf 63}, 034008 (2001)
  [hep-ph/0008207];
  E.~Nakano and T.~Tatsumi,
  Phys.\ Rev.\ D {\bf 71}, 114006 (2005)
  [hep-ph/0411350].

\bibitem{Basar:2008im} 
  G.~Basar and G.~V.~Dunne,
  Phys.\ Rev.\ Lett.\  {\bf 100}, 200404 (2008)
  [arXiv:0803.1501 [hep-th]];
  G.~Basar and G.~V.~Dunne,
  Phys.\ Rev.\ D {\bf 78}, 065022 (2008)
  [arXiv:0806.2659 [hep-th]];
G.~Basar, G.~V.~Dunne and M.~Thies,
  Phys.\ Rev.\ D {\bf 79}, 105012 (2009)
  [arXiv:0903.1868 [hep-th]].  

\bibitem{Nickel:2009ke} 
  D.~Nickel,
  Phys.\ Rev.\ Lett.\  {\bf 103}, 072301 (2009)
  [arXiv:0902.1778 [hep-ph]];
  D.~Nickel,
  Phys.\ Rev.\ D {\bf 80}, 074025 (2009)
  [arXiv:0906.5295 [hep-ph]];
  S.~Carignano, D.~Nickel and M.~Buballa,
  Phys.\ Rev.\ D {\bf 82}, 054009 (2010)
  [arXiv:1007.1397 [hep-ph]].

  \bibitem{Kojo:2009ha} 
  T.~Kojo, Y.~Hidaka, L.~McLerran and R.~D.~Pisarski,
  Nucl.\ Phys.\ A {\bf 843}, 37 (2010)
  [arXiv:0912.3800 [hep-ph]].

  \bibitem{Basar:2010zd} 
  G.~Basar, G.~V.~Dunne and D.~E.~Kharzeev,
  Phys.\ Rev.\ Lett.\  {\bf 104}, 232301 (2010)
  [arXiv:1003.3464 [hep-ph]];
  E.~J.~Ferrer, V.~de la Incera and A.~Sanchez,
  Acta Phys.\ Polon.\ Supp.\  {\bf 5}, 679 (2012)
  [arXiv:1205.4492 [nucl-th]].

\bibitem{Buballa:2014tba} 
 For a review, see e.g.,
  M.~Buballa and S.~Carignano,
  arXiv:1406.1367 [hep-ph].

\bibitem{Fulde:1964zz} 
  P.~Fulde and R.~A.~Ferrell,
  Phys.\ Rev.\  {\bf 135}, A550 (1964);
  A.~I.~larkin and Y.~N.~Ovchinnikov,
  Zh.\ Eksp.\ Teor.\ Fiz.\  {\bf 47}, 1136 (1964)
  [Sov.\ Phys.\ JETP {\bf 20}, 762 (1965)].

\bibitem{Polyakov:1978vu} 
  A.~M.~Polyakov,
  Phys.\ Lett.\ B {\bf 72}, 477 (1978);  
  B.~Svetitsky and L.~G.~Yaffe,
  Nucl.\ Phys.\ B {\bf 210}, 423 (1982).  

\bibitem{Cheng:2007jq} 
  M.~Cheng, N.~H.~Christ, S.~Datta, J.~van der Heide, C.~Jung, F.~Karsch, O.~Kaczmarek and E.~Laermann {\it et al.},
  Phys.\ Rev.\ D {\bf 77}, 014511 (2008)
  [arXiv:0710.0354 [hep-lat]];      
   A.~Bazavov, T.~Bhattacharya, M.~Cheng, N.~H.~Christ, C.~DeTar, S.~Ejiri, S.~Gottlieb and R.~Gupta {\it et al.},
  Phys.\ Rev.\ D {\bf 80}, 014504 (2009)
  [arXiv:0903.4379 [hep-lat]].  

\bibitem{Hatta:2003ga} 
  Y.~Hatta and K.~Fukushima,
  Phys.\ Rev.\ D {\bf 69}, 097502 (2004)
  [hep-ph/0307068].

\bibitem{Fukushima:2003fw} 
  K.~Fukushima,
  Phys.\ Lett.\ B {\bf 591}, 277 (2004)
  [hep-ph/0310121].

\bibitem{Yamamoto:2014lia} 
  A.~Yamamoto,
  Phys.\ Rev.\ Lett.\  {\bf 112}, 162002 (2014)
  [arXiv:1402.3049 [hep-lat]].

\bibitem{Dumitru:2005ng} 
  A.~Dumitru, R.~D.~Pisarski and D.~Zschiesche,
  Phys.\ Rev.\ D {\bf 72}, 065008 (2005)
  [hep-ph/0505256].

\bibitem{Roberge:1986mm} 
  A.~Roberge and N.~Weiss,
  Nucl.\ Phys.\ B {\bf 275}, 734 (1986).

\bibitem{Bonati:2013tt} 
  C.~Bonati, M.~D'Elia, H.~Panagopoulos and E.~Vicari,
  Phys.\ Rev.\ Lett.\  {\bf 110}, no. 25, 252003 (2013)
  [arXiv:1301.7640 [hep-lat]];
  M.~D'Elia and F.~Negro,
  Phys.\ Rev.\ D {\bf 88}, no. 3, 034503 (2013)
  [arXiv:1306.2919 [hep-lat]].

\bibitem{Huang:1990jf} 
  S.~Huang, J.~Potvin, C.~Rebbi and S.~Sanielevici,
  Phys.\ Rev.\ D {\bf 42}, 2864 (1990);
  {\it ibid}. {\bf 43}, 2056 (1991);
  Y.~Iwasaki, K.~Kanaya, L.~Karkkainen, K.~Rummukainen and T.~Yoshie,
  Phys.\ Rev.\ D {\bf 49}, 3540 (1994) [hep-lat/9309003];
  A.~Gopie and M.~C.~Ogilvie,
  Phys.\ Rev.\ D {\bf 59}, 034009 (1999) [hep-lat/9803005].

\bibitem{Yamamoto:2013zwa} 
  A.~Yamamoto,
  arXiv:1405.6665 [hep-lat].

\bibitem{Takahashi:2002bw} 
  T.~T.~Takahashi, H.~Suganuma, Y.~Nemoto and H.~Matsufuru,
  Phys.\ Rev.\ D {\bf 65}, 114509 (2002)
  [hep-lat/0204011].

\bibitem{McLerran:1980pk} 
  L.~D.~McLerran and B.~Svetitsky,
  Phys.\ Lett.\ B {\bf 98}, 195 (1981);
  Phys.\ Rev.\ D {\bf 24}, 450 (1981);
  S.~Nadkarni,
  Phys.\ Rev.\ D {\bf 33}, 3738 (1986).

\end{thebibliography}

\end{document}